\documentclass[10pt,conference]{IEEEtran} %,compsocconf,letterpaper
\IEEEoverridecommandlockouts
\usepackage{cite}
\usepackage{amsmath,amssymb,amsfonts}
\usepackage{graphicx}
\usepackage{textcomp}
\usepackage{xcolor}
\usepackage{caption}
\usepackage{subcaption}
\usepackage{tikz}
\usepackage{mathtools}
\usepackage{xcolor}

\usepackage{acronym}
\usepackage[binary-units=true]{siunitx}
\usepackage{circledsteps}
\usepackage{multicol}
\usepackage{algorithm}
\usepackage{algpseudocode}
\usepackage{amsmath}
\usepackage[shortlabels]{enumitem}

\usepackage{hyperref}
\usepackage{cleveref}

\def\BibTeX{{\rm B\kern-.05em{\sc i\kern-.025em b}\kern-.08em
    T\kern-.1667em\lower.7ex\hbox{E}\kern-.125emX}}
\begin{document}

\title{Wireless Large Object Transmission Under Safety Constraints (LOTUS)}

\author{
	\IEEEauthorblockN{Alex Bendrick, Daniel Tappe, Rolf Ernst}
	\IEEEauthorblockA{\textit{Institute of Computer and Network Engineering}, \textit{Technische Universit\"at Braunschweig}, Germany \\
		\{bendrick, tappe, ernst\}@ida.ing.tu-bs.de}
}

\maketitle

% circles with arbitrary chars
\newcommand*\circled[1]{\tikz[baseline=(char.base)]{
            \node[shape=circle,draw,inner sep=2pt] (char) {#1};}}
            
% boxes with arbitrary chars
\newcommand*\boxd[2]{\tikz[baseline=(char.base)]{
            \node[shape=rectangle,draw,inner sep=2pt,fill=#2] (char) {#1};}}
            
% argmax
%\DeclareMathOperator*{\argmax}{arg\,max}
%\DeclareMathOperator*{\argmin}{arg\,min}
\newcommand{\argmax}[1]{\underset{#1}{\operatorname{arg}\,\operatorname{max}}\;}

%% intervals
%\NewDocumentCommand{\INTERVALINNARDS}{ m m }{
%    #1 {,} #2
%}
%\NewDocumentCommand{\interval}{ s m >{\SplitArgument{1}{,}}m m o }{
%    \IfBooleanTF{#1}{
%        \left#2 \INTERVALINNARDS #3 \right#4
%    }{
%        \IfValueTF{#5}{
%            #5{#2} \INTERVALINNARDS #3 #5{#4}
%        }{
%            #2 \INTERVALINNARDS #3 #4
%        }
%    }
%}

% Highlight notes in the document
\newenvironment{notes}{\color{blue}}{}
\newenvironment{notes_adaptations}{\color{orange}}{}
\newenvironment{notes_new}{\color{green}}{}

% citation needed command
\newcommand{\citeme}{\textbf{\textcolor{red}{[citation needed]}}} % Highligh a reminder to add a citation
\newcommand{\refme}{\textbf{\textcolor{red}{[reference needed]}}} % Highligh a reminder to add a reference (figure, table, etc)

% math ceil and floor commands
\DeclarePairedDelimiter\ceil{\lceil}{\rceil}
\DeclarePairedDelimiter\floor{\lfloor}{\rfloor}

\newcommand{\rom}[1]{\MakeUppercase{\romannumeral #1}}

% TODOs
\definecolor{dkgreen}{rgb}{0,0.4,0}
\definecolor{mauve}{rgb}{0.58,0,0.82}
\definecolor{red}{rgb}{255,0,0}
\definecolor{gruen_pp}{RGB}{146,208,80}
\definecolor{blau_pp}{RGB}{255,192,0}%{91,155,213}
\definecolor{yellowish_pp}{RGB}{0,255,255}%{255,217,102}
\definecolor{purple_pp}{RGB}{255,1,255}%{255,217,102}
\def\todoColorAB{\color{mauve}}
\def\todoColorDT{\color{purple}}
\def\todoColorDS{\color{orange}}
\def\todoColorNS{\color{green}}
\def\todoColorXX{\color{red}}
\newcommand{\todo}[2]{{\csname todoColor#1\endcsname ToDo[#1]: #2}}

\acrodef{DDS}{Data Distribution Service}
\acrodef{RTPS}{Real-Time Publish-Subscribe}
\acrodef{UDP}{User Datagram Protocol}
\acrodef{TCP}{Transmission Control Protocol}
\acrodef{MAC}{Media Access Control}
\acrodef{CSMA/CA}{Carrier Sense Multiple Access with Collision Avoidance}
\acrodef{CW}{Contention Window}
\acrodef{SPS}{Semi-Persistent Scheduling}
\acrodef{BER}{Bit Error Rate}
\acrodef{RB}{Resource Block}
\acrodef{FDMA}{Frequency-Division Multiple Access}
\acrodef{OFDMA}{Orthogonal Frequency-Division Multiple Access}
\acrodef{MIMO}{Multiple Input, Multiple Output}
\acrodef{NC}{Network Coding}
\acrodef{OTA}{Over-the-Air}

\acrodef{DSRC}{Dedicated Short Range Communication}
\acrodef{WAVE}{Wireless Access in Vehicular Environments}
\acrodef{LTE}{Long-Term Evolution}

\acrodef{SDN}{Software-Defined Network}

\acrodef{W2RP}{Wireless Reliable Real-Time Protocol}
\acrodef{E-W2RP}{Enhanced-W2RP}
\acrodef{WiMEP}{Wireless Multicast Error Protection Protocol}

\acrodef{BER}[BER]{Bit Error Rate}
\acrodef{FER}[FER]{Frame Error Rate}
\acrodef{FEC}{Forward Error Correction}
\acrodef{BEC}{Backward Error Correction}
\acrodef{SNR}{signal to noise ratio}

\acrodef{SP}[$T_S$]{sample period}
\acrodef{SS}[$S_S$]{sample size}
\acrodef{SF}[$S_f$]{fragment size}
\acrodef{SFR}[$S_\text{fr}$]{frame size}
\acrodef{SD}[$D_S$]{sample deadline}
\acrodef{aat}[$\overline{t}_a$]{average arbitration time}

\acrodef{TO}[$t_{retx}$]{retransmission timeout interval}
\acrodef{TCFG}[$t_{cfg}$]{reconfiguration time}
\acrodef{PH}[$T_{HB}$]{heartbeat period}

\acrodef{V2X}{Vehicle-to-Everything}
\acrodef{VANET}{Vehicular Ad-hoc Network}
\acrodef{C-V2X}{Cellular V2X}
\acrodef{URLLC}{Ultra-Reliable and Low Latency Communications}
\acrodef{NR}{New Radio}
\acrodef{PDR}[$PDR_i$]{packet delivery rate}
\acrodef{QoS}{Quality of Service}
\acrodef{ROS2}{Robot Operating System 2}
\acrodef{A2D2}{Audi Autonomous Driving Dataset}
\acrodef{DVS}{Digital Vision Security}

\acrodef{AUTOSAR}{Automotive Open System Architecture}
\acrodef{RM}{Resource Management}
\acrodef{NMU}{Network Management Unit}
\acrodef{aRM}{application layer RM}
\acrodef{nRM}{network layer RM}
\acrodef{RRM}{Radio Resource Management}

\acrodef{AV}{autonomous vehicle}
\acrodef{MV}{manually controlled vehicle}
\acrodef{IN}{infrastructure node}

\acrodef{TSN}{Time-Sensitive Networking}

\acrodef{HB}{heartbeat}

\acrodef{GNSS}{Global Navigation Satellite System}
\acrodef{PTP}{Precision Time Protocol}

\acrodef{CAM}{Cooperative Awareness Message}
\acrodef{DENM}{Distributed Environmental Notification Message}
\acrodef{BSM}{Basic Safety Message}

\acrodef{RoI}{Region of Interest}
\acrodefplural{RoI}[RoIs]{Regions of Interest}

\acrodef{3GPP}{3rd Generation Partnership Project}

\acrodef{N}[$N_i$]{nodes}
\acrodef{C}[$C_i$]{context}
\acrodef{R}[$R_i$]{parameter set}

\acrodef{SOTIF}{safety of the intended function}

\acrodef{GE}{Gilbert Elliot}

\acrodef{AD}{automated driving}
\acrodef{AP}{access point}
\acrodef{AVP}{automated valet parking}

\acrodef{ARQ}{automatic repeat request}
\acrodef{HARQ}{hybrid automatic repeat request}
\acrodef{MCS}{modulation and coding scheme}

\acrodef{eMBB}{enhanced Mobile Broadband}

\acrodef{MBB}{make-before-break}
\acrodef{ROS}{Robot Operating System}

\begin{abstract}

Future autonomous mobile systems will greatly benefit from cooperation and real-time sensor data exchange using V2X communication.
In such applications, wireless communication has to cope with stringent real-time and safety constraints, a huge challenge given the inherently lossy wireless communication with highly dynamic channel and error conditions.
To meet the safety and real-time goals, the use of state-of-the-art (5G and 802.11) V2X technologies focuses on reliable exchange of small data objects, as in URLLC.
In contrast, reliable low-latency exchange of large data, such as camera frames, has received little attention, despite its predicted benefits in safe perception and cooperation. 
The LOTUS project, outlined in this paper, exploits the specific properties of large application data objects to develop novel, application-aware mechanisms for low-latency reliable large data exchange on top of existing and future V2X technologies.
Evaluation with statistical analysis, simulation, and physical prototypes demonstrates the feasibility of low-latency large data object exchange at unprecedented levels of reliability.

\end{abstract}

\section{Introduction}
\label{intro}

The project Large Object Transmission Under Safety Constraints (LOTUS) addresses challenges in the communication of mobile systems, like \ac{V2X} or mobile robots.
Numerous roadmaps and surveys within the scope of wireless (\ac{V2X}) communication envision an increasing amount of demanding services.
Examples are cooperative perception and remote driving services, which are crucial to further enhance automated mobility (SAE Level 4+).
These applications require the reliable exchange of large data objects (camera images, LIDAR point clouds, occupancy grid maps, object lists) in real-time.

Wireless communication is inherently lossy due to fading effects like reflections, shadowing, and Doppler shift, impacting the signal-to-noise ratio (SNR) and causing bit errors.
Additionally, vehicle mobility necessitates constant roaming - and thus frequent handovers - between \acp{AP}, disrupting data transmission and compromising reliability.
Resilient and continuous application operation requires robustness against these common failure modes.

Current wireless and \ac{V2X} communication standardization efforts mainly involve the exchange of small data for sharing status and intent information.
Hence, reliability and low-latency mechanisms are optimized for small, single-packet data.
Such mechanisms are not able to consider a packet as part of a data object’s context, and thus insufficiently address the constraints of large data object transmissions.
Consequently, novel solutions for a reliable transmission of large data objects under real-time constraints are needed.

For this purpose, we propose application-centric solutions that overcome the issues of state-of-the-art \ac{V2X}/wireless technologies.
Specifically, application-centric approaches allow for utilizing application-information/knowledge for sophisticated coordination, exploiting sample-level deadlines and inherent slack for robustness against various error sources.

\section{Towards Reliable Sample Transmission}
\label{intro}

Sensor data streaming requires high data rates and reliability guarantees.
While dedicated \ac{V2X} standards (802.11p/bd or \ac{C-V2X}) offer reasonable robustness, they are primarily intended for transmitting small messages (\ac{CAM}, \ac{BSM}, \ac{DENM}, ...).
As such, with effective data rates smaller than \SI[per-mode=repeated-symbol]{30}{\mega\bit\per\second} large data streaming is not viable.
Current commodity wireless standards (802.11ax/be and cellular 5G) offer increasingly high data rates exceeding \SI[per-mode=repeated-symbol]{1}{\giga\bit\per\second}.
To protect transmission against unavoidable packet loss packet-level \ac{ARQ} and \ac{HARQ} mechanisms are employed to retransmit missing packets.
However, theses packet-level \ac{BEC} schemes become ineffective when transmitting large fragmented samples.
Consequently, there is no existing wireless standard allowing for reliable sample streaming.

\subsubsection*{\textbf{Paper 1}}
\ac{W2RP} \cite{peeck_middleware_2021} is the first protocol that specifically addresses the reliable and timely exchange of large, fragmented samples in \ac{V2X} environments.
In contrast to the small, single-packet messages exchanged via \ac{V2X} technologies, multi-packet samples can be exchanged within an extended application-level deadline.
By deviating from packet-level considerations and taking advantage of those sample-level larger deadlines, \ac{W2RP} maximizes the efficiency of \ac{BEC}, which improves reliability and minimizes the allocated resources of large \ac{V2X} data streams.
Simulative results highlighted significant improvements over packet-level MAC-layer \ac{BEC}, that only managed sample transmissions in channels with low ($<$ \SI{10}{\percent}) whereas \ac{W2RP} ensured reliable sample transmission up to error rates of \SI{50}{\percent}.
\ac{W2RP} is based on the popular Data Distribution Service (DDS), which is the default communication middleware of the \ac{ROS} 2 and is also part of the AUTOSAR communication specification.
This makes W2RP easily accessible for robotic and automotive applications.

\section{Enhanced Robustness}
\label{robust}

Despite modern wireless technologies such as 802.11ax/be (WiFi6/7) and 5G cellular boasting higher data rates as well as higher claimed reliability, reliable large data exchange is still challenging.
Typically, there are two dedicated modes:
High data rate (cellular \ac{eMBB} and commodity WiFi) modes typically utilize less robust coding schemes  and technologies like mmWave that are more susceptible to various fading effects to boost data rates.
However, the existing reliability mechanisms are inadequate to meet the stringent constraints imposed on the data exchange by safety-critical application.
As a result, these two high data rate modes are no viable solution to enable cooperative perception and similar applications requiring hard real-time streaming of large data.
(Ultra) high reliability mechanisms (\ac{URLLC} and wireless \ac{TSN}), on the other hand, are only applicable for small data.
A combination of high reliability and high data rates is not intended by the standards.
Consequently, research on dedicated robustness mechanisms is needed.
As even current high reliability mechanisms from wireless standards are insufficient for large sample streaming, \ac{W2RP} was further developed in multiple directions, focused on improving robustness under various scenarios and channel conditions.

\subsubsection*{\textbf{Paper 2}}
First, unicast data dissemination in cooperative perception and similar applications is inefficient as data may be needed by multiple vehicles/nodes.
Hence, \ac{W2RP} has been extended to support multicast \cite{bendrick_error_2023}.
For this purpose, bundling of \ac{BEC} for receivers with similar error patterns and means to prioritize receivers based on arbitrary conditions are introduced.
The effectiveness of the multicast mechanism for improving robustness has been demonstrated using simulations\footnote{https://github.com/IDA-TUBS/IDAWirelessSimulator} as well as the IDA's physical demonstrator setup.

\subsubsection*{\textbf{Paper 3}}
First works on \ac{W2RP} assumed uniformly distributed (bit) errors.
However, real-world scenarios can be subject to more sophisticated error conditions.
For example, burst errors can occur that put additional strain on the \ac{BEC} mechanism of \ac{W2RP}.
By exploiting application properties and constraints (high sampling rates and sample deadlines) to enabling overlapping yet more robust retransmission robustness to burst errors was drastically improved  \cite{bendrick_hard_2023}.

\subsubsection*{\textbf{Paper 4}}
Additionally, \ac{W2RP} has been extended by means of a decentralized parameter selection \cite{peeck_enabling_2023} to improve effectiveness in shared channels.

\subsubsection*{\textbf{Paper 5}}
Finally, an analysis of real-world data sets and perception pipelines gave valuable insights into what parts of data are actually required to be exchanged for cooperative perception purposes \cite{bendrick2024subsample}:
By offering two complementary, lossless and data-aware data optimization mechanisms, namely transmitting \acp{RoI} or incremental updates, \ac{W2RP} slack -- and thereby robustness to various errors -- can be increased.
Thereby, \ac{RoI} are especially useful for sensors mounted to mobile systems whereas incremental updates are most advantageous for statically mounted sensors, e.g., infrastructure cameras. 
%For static infrastructure-mounted (camera) sensors it become apparent, that typically only parts of an frame change, as typically on a selection of traffic participants is actually moving.
%Therefore, only transmitting incremental updates would suffice for giving receivers a complete picture of their surroundings.
%Both mechanisms can significantly decrease the size of samples without loss in data quality, allowing for higher robustness to various (transient) error effects.
\section{Coordination by means of Resource Management}
\label{rm}

Due to its shared nature, the wireless channel is used by various applications of different criticalities and with different requirements and constraints.
This makes timing guarantees impractical, as long as no coordination is used.
Existing (cellular) wireless standards and scientific literature propose various means for arbitrating resource allocation.
Mostly, these comprise \ac{MAC}, \ac{RRM} and \ac{SDN} techniques.
Critically, however, these mechanisms typically continue to focus exclusively on the packet level and do not allow for consideration of large samples, thus limiting their effectiveness in ensuring reliable sample exchange.

\subsubsection*{\textbf{Paper 6}}
A centralized application-centric \ac{RM} approach that is based on \ac{SDN} principles (communication divided into control and data plane) has been developed to address the coordination of various wireless applications \cite{ernst_application-centric_2023}.
A hierarchical architecture is extending the \ac{SDN} network control layer by an application control layer.
It supports application-aware coordination across network segments.
Thereby, the \ac{RM} takes into account both application requirements and constraints as well as network properties.

\subsubsection*{\textbf{Paper 7}}
Due to the safety-critical nature of applications, any coordination and reconfiguration of applications under such circumstances must adhere to stringent timing and safety constraints itself to prevent disruption of application service.
This aspect has not been addressed by the application-centric \ac{RM} previously.
Thus, a dedicated reconfiguration protocol has been developed  \cite{bendrick2024rm}. 
The combination of \ac{HB}-based connection monitoring, protection against inherent message loss, node-level fail-silent behavior and synchronized mode changes the \ac{RM} ensures safety by enforcing consistent application (and network) modes across the wireless network.

\subsubsection*{\textbf{Paper 8}}
The coordination principles have further been extended with respect to letting applications on a single node share resources ('shared slack') to enable more efficient resource utilization and to better cope with dynamic \ac{BEC} needs in wireless channels \cite{bendrick2024sharedslack}.

\section{Continuous Connectivity in Handover Scenarios}
\label{handover}

Handover situations, such as present in infrastructure-supported cooperative perception use cases such as \ac{AVP} or camera-equipped intersections, are especially taxing with respect to enable reliable and timely data exchange.
The mobile nature of vehicles inherently leads to the need for roaming across multiple \aclp{AP} (\acp{AP}), as shadowing by stationary (e.g., walls, vegetation, ...) and dynamic (e.g., other vehicles, ...) obstacles will occur frequently, especially in case high-data rate technologies are used.
In current cell-centric networks, handovers typically result in intermediate connection loss, making reliable sample exchange impossible.
While the cellular 5G standard tries to improve upon this by employing dual connectivity and a \ac{MBB} approach that enables 'lossless' handovers (at least for \ac{URLLC}), there are still no guarantees and also limited determinism with respect to handover timing (1-50 \SI{}{\milli\second}).
Notably, 'normal' non-\ac{URLLC} take even longer ($>$ \SI{50}{\milli\second}) and offer even less determinism making continuous sample exchange impractical using state-of-the-art wireless standards.
Consequently, novel mechanisms to address this issue and enable continuous streaming of large data are required.

\subsubsection*{\textbf{Paper 9}}
For this purpose a continuous connectivity approach has been developed \cite{tappe2024handover}.
It utilizes multi-connectivity (cf. cell-free architecture), where all connections are monitored at the control layer, but only one connection is used to stream the data.
A low latency \ac{TSN} network connects the \acp{AP} and reroutes the data to another \ac{AP} in case of a connection loss, exploiting the slack of the data plane.
This way, the total data rate does not increase beyond that of a single stream. 
As the backbone network has a fixed topology, safe alternative routes can be determined in advance, thereby taking this route determination out of the critical path.
Consequently, the reconfiguration time is bounded from above by the deterministic connection loss detection and switching of routes.
Physical experiments showed deterministic handover within less than \SI{10}{\milli\second} being achievable, thereby ensuring reliable sample transmission even in handover situations.
\section*{Further Research}

For further publications and updates please check the LOTUS website (\href{LOTUS}{\textcolor{blue}{https://ida-tubs.github.io/lotus/}}).
In-depth information on each topic can be found within the respective papers.
The list of these papers can be found below.

\renewcommand{\refname}{Papers}
% references
\bibliographystyle{IEEEtran}
\bibliography{bibliography}

% Generated by IEEEtran.bst, version: 1.14 (2015/08/26)
\begin{thebibliography}{1}
\providecommand{\url}[1]{#1}
\csname url@samestyle\endcsname
\providecommand{\newblock}{\relax}
\providecommand{\bibinfo}[2]{#2}
\providecommand{\BIBentrySTDinterwordspacing}{\spaceskip=0pt\relax}
\providecommand{\BIBentryALTinterwordstretchfactor}{4}
\providecommand{\BIBentryALTinterwordspacing}{\spaceskip=\fontdimen2\font plus
\BIBentryALTinterwordstretchfactor\fontdimen3\font minus
  \fontdimen4\font\relax}
\providecommand{\BIBforeignlanguage}[2]{{%
\expandafter\ifx\csname l@#1\endcsname\relax
\typeout{** WARNING: IEEEtran.bst: No hyphenation pattern has been}%
\typeout{** loaded for the language `#1'. Using the pattern for}%
\typeout{** the default language instead.}%
\else
\language=\csname l@#1\endcsname
\fi
#2}}
\providecommand{\BIBdecl}{\relax}
\BIBdecl

\bibitem{peeck_middleware_2021}
J.~Peeck, M.~Möstl, T.~Ishigooka, and R.~Ernst, ``A {Middleware} {Protocol}
  for {Time}-{Critical} {Wireless} {Communication} of {Large} {Data}
  {Samples},'' in \emph{2021 IEEE Real-Time Systems Symposium (RTSS)}, Dec.
  2021, pp. 1--13.

\bibitem{bendrick_error_2023}
A.~Bendrick, J.~Peeck, and R.~Ernst, ``An {Error} {Protection} {Protocol} for
  the {Multicast} {Transmission} of {Data} {Samples} in {V2X} {Applications},''
  \emph{ACM Transactions on Cyber-Physical Systems}, Aug. 2023.

\bibitem{bendrick_hard_2023}
A.~Bendrick and R.~Ernst, ``{Hard Real-Time Streaming of Large Data Objects
  with Overlapping Backward Error Correction},'' in \emph{IECON 2023- 49th
  Annual Conference of the IEEE Industrial Electronics Society}, 2023, pp.
  1--8.

\bibitem{peeck_enabling_2023}
J.~Peeck and R.~Ernst, ``Enabling multi-link data transmission for
  collaborative sensing in open road scenarios,'' in \emph{Proceedings of the
  31st International Conference on Real-Time Networks and Systems (RTNS 2023)},
  Jun. 2023, pp. 76--86.

\bibitem{bendrick2024subsample}
A.~Bendrick, N.~Sperling, and R.~Ernst, ``{Large Data Transfer Optimization for
  Improved Robustness in Real-Time V2X-Communication},'' \emph{{IEEE
  Transactions on Computer-Aided Design of Integrated Circuits and Systems}},
  2024.

\bibitem{ernst_application-centric_2023}
R.~Ernst, D.~Stöhrmann, A.~Bendrick, and A.~Kostrzewa, ``Application-centric
  {Network} {Management} - {Addressing} {Safety} and {Real}-time in {V2X}
  {Applications},'' \emph{ACM Transactions on Embedded Computing Systems},
  vol.~22, no.~2, pp. 20:1--20:25, Jan. 2023, number: 2.

\bibitem{bendrick2024rm}
A.~Bendrick, D.~Tappe, D.~St\"ohrmann, and R.~Ernst, ``{Synchronized Lossfree
  Reconfiguration of Safety-critical V2X Streaming Applications},'' \emph{{IEEE
  Transactions on Vehicular Technology}}, 2024.

\bibitem{bendrick2024sharedslack}
A.~Bendrick, D.~Tappe, and R.~Ernst, ``{Ultra Reliable Hard Real-Time V2X
  Streaming with Shared Slack Budgeting},'' in \emph{{2024 IEEE Intelligent
  Vehicles Symposium (IV)}}, 2024.

\bibitem{tappe2024handover}
D.~Tappe, A.~Bendrick, and R.~Ernst, ``{Continuous multi-access communication
  for high-resolution low-latency V2X sensor streaming},'' in \emph{{2024 IEEE
  Intelligent Vehicles Symposium (IV)}}, 2024.

\end{thebibliography}

\typeout{get arXiv to do 4 passes: Label(s) may have changed. Rerun}
 	
\end{document}